\documentstyle{article}

\catcode`\@=11
\def\@citex[#1]#2{%
\if@filesw \immediate \write \@auxout {\string \citation {#2}}\fi
\@tempcntb\m@ne \let\@h@ld\relax \def\@citea{}%
\@cite{%
  \@for \@citeb:=#2\do {%
    \@ifundefined {b@\@citeb}%
      {\@h@ld\@citea\@tempcntb\m@ne{\bf ?}%
      \@warning {Citation `\@citeb ' on page \thepage \space undefined}}%
      {\@tempcnta\@tempcntb \advance\@tempcnta\@ne%
      \@tempcntb\number\csname b@\@citeb \endcsname \relax%
      \ifnum\@tempcnta=\@tempcntb 
        \ifx\@h@ld\relax%
          \edef \@h@ld{\@citea\csname b@\@citeb\endcsname}%
        \else%
          \edef\@h@ld{\ifmmode{-}\else--\fi\csname b@\@citeb\endcsname}%
        \fi%
      \else
        \@h@ld\@citea\csname b@\@citeb \endcsname%
        \let\@h@ld\relax%
      \fi}%
    \def\@citea{,\penalty\@highpenalty\,}%
  }\@h@ld
}{#1}}

\def\@citeb#1#2{{[#1]\if@tempswa , #2\fi}}
\def\@citeu#1#2{{$^{#1}$\if@tempswa , #2\fi }}
\def\@citep#1#2{{#1\if@tempswa , #2\fi}}

\def\bcites{         
        \catcode`\@=11
        \let\@cite=\@citeb
        \catcode`\@=12
}

\def\upcites{         
        \catcode`\@=11
        \let\@cite=\@citeu
        \catcode`\@=12
}

\def\plaincites{      
        \catcode`\@=11
        \let\@cite=\@citep
        \catcode`\@=12
}

\newcount\hour
\newcount\minute
\newtoks\amorpm
\hour=\time\divide\hour by 60
\minute=\time{\multiply\hour by 60 \global\advance\minute by-\hour}
\edef\standardtime{{\ifnum\hour<12 \global\amorpm={am}%
        \else\global\amorpm={pm}\advance\hour by-12 \fi
        \ifnum\hour=0 \hour=12 \fi
        \number\hour:\ifnum\minute<10 0\fi\number\minute\the\amorpm}}
\edef\militarytime{\number\hour:\ifnum\minute<10 0\fi\number\minute}

\def\draftlabel#1{{\@bsphack\if@filesw {\let\thepage\relax
   \xdef\@gtempa{\write\@auxout{\string
      \newlabel{#1}{{\@currentlabel}{\thepage}}}}}\@gtempa
   \if@nobreak \ifvmode\nobreak\fi\fi\fi\@esphack}
        \gdef\@eqnlabel{#1}}
\def\@eqnlabel{}
\def\@vacuum{}
\def\marginnote#1{}
\def\draftmarginnote#1{\marginpar{\raggedright\scriptsize\tt#1}}
\def\draft{
        \pagestyle{plain}
        \overfullrule=2pt
        \oddsidemargin -.5truein
        \def\@oddhead{\sl \phantom{\today\quad\militarytime} \hfil
        \smash{\Large\sl DRAFT} \hfil \today\quad\militarytime}
        \let\@evenhead\@oddhead
        \let\label=\draftlabel
        \let\marginnote=\draftmarginnote
        \def\ps@empty{\let\@mkboth\@gobbletwo
        \def\@oddfoot{\hfil \smash{\Large\sl DRAFT} \hfil}
        \let\@evenfoot\@oddhead}
        \def\@eqnnum{(\theequation)\rlap{\kern\marginparsep\tt\@eqnlabel}%
        \global\let\@eqnlabel\@vacuum}  }

\def\blackfonts{
        \font\blackboard=msbm10 scaled\magstep1
        \font\blackboards=msbm8
        \font\blackboardss=msbm6
}

\def\prep{         
        \catcode`\@=11
        \input art10.sty
        \catcode`\@=12
        
        \let\small\null
        \def\blackfonts{
                \font\blackboard=msbm10
                \font\blackboards=msbm7
                \font\blackboardss=msbm5
        }
        \let\sl\it
        \twocolumn
        \sloppy
        \voffset=-2.54truecm
        \hoffset=-2.54truecm
        \flushbottom
        \parindent 1em
        \leftmargini 2em
        \leftmarginv .5em
        \leftmarginvi .5em
        \marginparwidth 48pt
        \marginparsep 10pt
        \setlength{\columnsep}{2truecm}
        \setlength{\textwidth}{25.4truecm}
        \setlength{\textheight}{17truecm}
        \baselineskip=16pt
        \oddsidemargin .18truein
        \evensidemargin .17truein
}

\def\eqalign#1{\null\,\vcenter{\openup\jot\m@th
  \ialign{\strut\hfil$\displaystyle{##}$&$\displaystyle{{}##}$\hfil
      \crcr#1\crcr}}\,}
\def\eqalignno#1{\displ@y \tabskip\centering
  \halign to\displaywidth{\hfil$\@lign\displaystyle{##}$\tabskip\z@skip
    &$\@lign\displaystyle{{}##}$\hfil\tabskip\centering
    &\llap{$\@lign##$}\tabskip\z@skip\crcr
    #1\crcr}}

\def\section{\@startsection {section}{1}{\z@}{3.ex plus 1ex minus
 .2ex}{2.ex plus .2ex}{\large\bf}}
\def\subsection{\@startsection{subsection}{2}{\z@}{2.75ex plus 1ex minus
 .2ex}{1.5ex plus .2ex}{\bf}}        

\def\appendix{{\newpage\section*{Appendix}}\let\appendix\section%
        {\setcounter{section}{0}
        \gdef\thesection{\Alph{section}}}\section}

\def\abstract{\if@twocolumn
\section*{Abstract}
\else 
\begin{center}
{\bf Abstract\vspace{-.5em}\vspace{0pt}}
\end{center}
\quotation
\fi}

\catcode`\@=12

\def\d{\partial}

\def\sqr#1#2{{\vcenter{\vbox{\hrule height.#2pt\hbox{\vrule width.#2pt 
height#1pt \kern#1pt \vrule width.#2pt}\hrule height.#2pt}}}}

\def\=d{\,{\buildrel\rm def\over =}\,}

\def\F{{\cal F}}

\def\i3p{\p32\int d^3p}

\def\As{A\hbox to 1pt{\hss /}}
\def\np4{\int d^4p_1\cdots d^4p_{n-1}\, }

\def\Tr{{\rm Tr}\, }

\def\nx4{\int d^4x_1\ldots d^4x_n\, }

\def\kon#1#2{\vbox{\halign{##&&##\cr
\lower4pt\hbox{$\scriptscriptstyle\vert$}\hrulefill &
\hrulefill\lower4pt\hbox{$\scriptscriptstyle\vert$}\cr $#1$&
$#2$\cr}}}

\def\konv#1#2#3{\hbox{\vrule height12pt depth-1pt}
\vbox{\hrule height12pt width#1cm depth-11.6pt}
\hbox{\vrule height6.5pt depth-0.5pt}
\vbox{\hrule height11pt width#2cm depth-10.6pt\kern5pt
      \hrule height6.5pt width#2cm depth-6.1pt}
\hbox{\vrule height12pt depth-1pt}
\vbox{\hrule height6.5pt width#3cm depth-6.1pt}
\hbox{\vrule height6.5pt depth-0.5pt}}
\def\konu#1#2#3{\hbox{\vrule height12pt depth-1pt}
\vbox{\hrule height1pt width#1cm depth-0.6pt}
\hbox{\vrule height12pt depth-6.5pt}
\vbox{\hrule height6pt width#2cm depth-5.6pt\kern5pt
      \hrule height1pt width#2cm depth-0.6pt}
\hbox{\vrule height12pt depth-6.5pt}
\vbox{\hrule height1pt width#3cm depth-0.6pt}
\hbox{\vrule height12pt depth-1pt}}

\def\konw#1#2#3{\hbox{\vrule height12pt depth-1pt}
\vbox{\hrule height12pt width#1cm depth-11.6pt}
\hbox{\vrule height6.5pt depth-0.5pt}
\vbox{\hrule height12pt width#2cm depth-11.6pt \kern5pt
      \hrule height6.5pt width#2cm depth-6.1pt}
\hbox{\vrule height6.5pt depth-0.5pt}
\vbox{\hrule height12pt width#3cm depth-11.6pt}
\hbox{\vrule height12pt depth-1pt}}

\def\i{{\rm int}}

\def\c{{\rm cl}}
\def\e{{\rm ext}}

\def\r{{\rm ret}}

\def\m3{{\mu_1\mu_2\mu_3}}

\def\co{{\rm Com}}

\def\p{{(+)}}

\def\be{\begin{equation}}       \def\eq{\begin{equation}}
\def\ee{\end{equation}}         \def\eqe{\end{equation}}

\def\bea{\begin{eqnarray}}      \def\eqa{\begin{eqnarray}}
\def\ena{\end{eqnarray}}        \def\eea{\end{eqnarray}}
                                \def\eqae{\end{eqnarray}}

\def\ba{\begin{array}}
\def\ea{\end{array}}
\def\unit{1 \hskip-.3em \raise2pt\hbox{$ \scriptstyle |$ } }

\def\c{\gamma} 
\def\d{\delta}
\def\e{\epsilon}           
\def\f{\phi}               
\def\g{\gamma}
   
\def\i{\iota}

\def\l{\lambda}
\def\m{\mu}
\def\n{\nu}
  
\def\p{\pi}                
\def\r{\rho}                                     
\def\s{\sigma}                                   
\def\t{\tau}

\def\D{\Delta}
\def\F{\Phi}
\def\G{\Gamma}

\def\L{\Lambda}

\def\cl{{\cal L}}

\def\co{{\cal O}}

\def\half{{1 \over 2}}

\def\bop#1{\setbox0=\hbox{$#1M$}\mkern1.5mu
        \vbox{\hrule height0pt depth.04\ht0
        \hbox{\vrule width.04\ht0 height.9\ht0 \kern.9\ht0
        \vrule width.04\ht0}\hrule height.04\ht0}\mkern1.5mu}
\def\Box{{\mathpalette\bop{}}}                        
\def\pa{\partial}                              

\def\>{\rangle} 

\def\<{\langle} 
\def\Dsl{D \hskip-.6em \raise1pt\hbox{$ / $ } }

\def\sl#1{\rlap{\hbox{$\mskip 1 mu /$}}#1}
\def\leftrightarrowfill{$\mathsurround=0pt \mathord\leftarrow \mkern-6mu
       \cleaders\hbox{$\mkern-2mu \mathord- \mkern-2mu$}\hfill
       \mkern-6mu \mathord\rightarrow$}
\def\dvec#1{\vbox{\ialign{##\crcr
       \leftrightarrowfill\crcr\noalign{\kern-1pt\nointerlineskip}
       $\hfil\displaystyle{#1}\hfil$\crcr}}}          
\def\hook#1{{\vrule height#1pt width0.4pt depth0pt}}
\def\leftrighthookfill#1{$\mathsurround=0pt \mathord\hook#1
       \hrulefill\mathord\hook#1$}
\def\underhook#1{\vtop{\ialign{##\crcr                 
       $\hfil\displaystyle{#1}\hfil$\crcr
       \noalign{\kern-1pt\nointerlineskip\vskip2pt}
       \leftrighthookfill5\crcr}}}
\def\smallunderhook#1{\vtop{\ialign{##\crcr      
       $\hfil\scriptstyle{#1}\hfil$\crcr
       \noalign{\kern-1pt\nointerlineskip\vskip2pt}
       \leftrighthookfill3\crcr}}}

\def\sfrac#1#2{{\vphantom1\smash{\lower.5ex\hbox{\small$#1$}}\over
       \vphantom1\smash{\raise.4ex\hbox{\small$#2$}}}} 
\def\bfrac#1#2{{\vphantom1\smash{\lower.5ex\hbox{$#1$}}\over
       \vphantom1\smash{\raise.3ex\hbox{$#2$}}}}      
\def\afrac#1#2{{\vphantom1\smash{\lower.5ex\hbox{$#1$}}\over#2}}  
\def\on#1#2{{\buildrel{\mkern2.5mu#1\mkern-2.5mu}\over{#2}}}
\def\ddt#1{\on{\hbox{\LARGE .\kern-2pt.}}#1}             
\def\tdt#1{\on{\hbox{\LARGE .\kern-2pt.\kern-2pt.}}#1}   

\def\boxes#1{
       \newcount\num
       \num=1
       \newdimen\downsy
       \downsy=-1.5ex
       \mskip-2.8mu
       \bo
       \loop
       \ifnum\num<#1
       \llap{\raise\num\downsy\hbox{$\bo$}}
       \advance\num by1
       \repeat}
\def\boxup#1#2{\newcount\numup
       \numup=#1
       \advance\numup by-1
       \newdimen\upsy
       \upsy=.75ex
       \mskip2.8mu
       \raise\numup\upsy\hbox{$#2$}}

\newskip\humongous \humongous=0pt plus 1000pt minus 1000pt
\def\caja{\mathsurround=0pt}
\def\eqalign#1{\,\vcenter{\openup2\jot \caja
       \ialign{\strut \hfil$\displaystyle{##}$&$
       \displaystyle{{}##}$\hfil\crcr#1\crcr}}\,}
\newif\ifdtup

\def\to{\rightarrow}

\def\1ov4{{1\over 4}}

\def\Tr{{\rm Tr}}

\def\pa{\partial}

\def\ddt{\dot{\t}}

\def\pa{\partial}

\def\nonu{\nonumber \\{}}
\def\half{{1 \over 2}}

\def\Ric{{\mbox{Ric}}}

\def\g0{g_{(0)}}

\newcommand{\sm}[1]{\mbox{\scriptsize #1}} 
\newcommand{\tnnn}[1]{\mbox{\tiny #1}} 

\def\d{\mbox{d}} 
\renewcommand{\theequation}{\thesection.\arabic{equation}}

\begin{document}

\large
\begin{flushright}
SPIN-2000/30\\
ITP-UU-00/32\\
PUTP-1970\\
{\tt hep-th/0011230}
\end{flushright}

\begin{center}
\vskip 1truecm
{\Large\bf
Gravity in Warped Compactifications and}\\
\vspace{.3cm}
{\Large\bf the Holographic Stress Tensor}\\
\vskip .8truecm

{\large\bf Sebastian de Haro, ${}^{\star\dagger}${}\footnote{e-mail: {\tt haro@phys.uu.nl}} Kostas Skenderis${}^{\ddagger}$\footnote{e-mail: {\tt kostas@feynman.princeton.edu}} and Sergey N. Solodukhin${}^{\star\star}${}\footnote{
e-mail: {\tt soloduk@theorie.physik.uni-muenchen.de}}}\\ 
\vskip 1truecm
${}^{\star}$ {\it Spinoza Institute, Utrecht University\\
Leuvenlaan 4, 3584 CE Utrecht, The Netherlands}
\vskip 1truemm
${}^{\dagger}$ {\it Institute for Theoretical Physics, Utrecht University\\
Princetonplein 5, 3584 CC Utrecht, The Netherlands}
\vskip 1truemm
${}^{\ddagger}$ {\it Physics Department, Princeton University \\
Princeton, NJ 08544, USA}
\vskip 1truemm
${}^{\star\star}$ {\it Theoretische Physik,
Ludwig-Maximilians Universitat, \\
Theresienstrasse 37,
D-80333, Munchen, Germany}

\end{center}
\vskip .5truecm
We study gravitational aspects of Brane-World scenarios.
We show that the bulk Einstein equations together with the 
junction  condition imply that the induced metric on the 
brane satisfies the full non-linear Einstein equations with a specific 
effective stress energy tensor. This result holds 
for any value of the bulk cosmological constant.
The analysis is done by either placing the brane close to infinity
or by considering the local geometry near the brane. 
In the case that the bulk spacetime is asymptotically AdS, we show that the 
effective stress energy tensor is equal to the sum
of the stress energy tensor of matter localized on the 
brane and of the holographic stress energy tensor appearing in 
the AdS/CFT duality. In addition, there are specific
higher-curvature corrections to Einstein's equations. 
We analyze in detail the case of asymptotically flat spacetime. 
We obtain asymptotic solutions of Einstein's
equations and show that the effective Newton's constant on the brane
depends on the position of the brane. 
 
\large

\newpage

\section{Warped Compactifications and AdS/CFT holography}

In the AdS/CFT correspondence the supergravity partition function is 
related to the generating functional of conformal field
theory (CFT) correlation functions as
\be\label{partf}
Z[\F]=\int_\f D\F\,\exp(iS[\F])=W_{\tnnn{CFT}}[\f],
\ee
where $\F$ denotes collectively all fields and $\f$ is a field parametrizing
the boundary condition of $\F$ at infinity. In the conformal field theory
the boundary fields $\f$ are interpreted as sources for CFT operators.
In particular, the metric at infinity, $g_{(0)}$, is considered as the 
source for the stress energy tensor of the dual CFT. The relation 
(\ref{partf})
suffers from divergences and has to be regularized and renormalized.

On the CFT side, there are UV divergences when operators come to 
coincident points. These correspond to IR divergences on the gravitational
side. To regulate the gravitational theory one may cut-off the 
asymptotically AdS spacetime at some radius $\r=\e$ near the boundary.
One can then compute all infrared divergences.
The renormalized theory is obtained by adding counterterms to 
cancel the infinities and then removing the cut-off.

One may wish, however, to consider situations where the infrared cutoff 
is kept finite instead of being sent to zero. This is the case in warped 
compactifications where the AdS spacetime is cut-off by the presence of 
a brane. In this case, (\ref{partf}) does not have any infrared divergences 
and so one does not need to add counterterms. 

In the cut-off spacetime, the induced metric at the boundary $\c$
corresponds to a normalizable mode and so one should integrate over it:
\be\label{1}
\int D\c_\e\int_{\c_\e}DG\exp(iS[G])=\int D\c_{\e}\,
W_{\tnnn{CFT}}[\c, \e],
\ee
Under these circumstances, gravity becomes 
dynamical on the brane, and the brane theory is a CFT coupled to 
dynamical gravity.

Consider a spacetime $M$ with a boundary $\pa M$. 
The action in (\ref{1}) is given by\footnote{
Our curvature conventions are as follows
$R_{ijk}{}^l=\pa_i \G_{jk}{}^l + \G_{ip}{}^l \G_{jk}{}^p - i
\leftrightarrow j$ and $R_{ij}=R_{ikj}{}^k$. With these conventions
the curvature of AdS comes out positive, but we will 
still use the terminology ``space of constant negative
curvature''. Notice also that we take
$\int \d^{d+1} x = \int \d^d x \int_0^\infty \d \r$
and the boundary is at $\r=0$.
The minus sign in front of the trace of the second 
fundamental form is correlated with the choice of having $\r=0$ in 
the lower end of the radial integration.}
\bea \label{action}
S[\F, G]&=&{1 \over 16 \p G_{d+1}}[\int_{M}\d^{d+1}x\, 
\sqrt{G}\, (R[G] + 2 \L) 
- \int_{\pa M} \d^d x\, \sqrt{\c}\, 2 K ] \nonu
&&+\int_{M}\d^{d+1}x\,\sqrt{G}\,  \cl^{\sm{bulk}}  
+\int_{\pa M} \d^d x\, \sqrt{\c}\, \cl^{\sm{bdry}}
\eea
where $\cl^{\sm{bulk}}$ denotes the Lagragian for bulk matter
and $\cl^{\sm{bdry}}$ the Lagrangian for matter
living on the boundary. Einstein's equations read\footnote{
The different signs in the right hand side of these two equations
is related to our conventions discussed in the previous 
footnote.}:
\bea\label{eom}
R_{\m\n}[G]-{1\over2}\,(R[G]+2\L)\,G_{\m\n} &=&
-8\p G_{d+1}\,T^{\sm{bulk}}_{\m\n}[G] \\
K_{ij}[\c]-\c_{ij}\,K[\c]&=&8\p G_{d+1}\,T_{ij}^{\sm{bdry}}[\c]. 
\label{junction}
\eea
These equations describe the case the bulk spacetime ends on the 
brane. This is in fact half of the Randall-Sundrum (RS) spacetime \cite{RS}.
In the RS scenario one glues on the other side of the brane an 
identical spacetime. Then the substitution
$$
K_{ij} \to \lim_{\delta \to 0}[K_{ij}(\r=\e+\delta) 
- K_{ij}(\r=\e-\delta)],
$$
in (\ref{junction})
yields the junction condition (see, for example, \cite{jct} for a
derivation). $\r=\e$ is the position of the brane.
In the RS context,
$K_{ij}(\r=\e+\delta)= - K_{ij}(\r=\e-\delta)$  due to the $Z_2$ symmetry, 
so the net effect is to get back (\ref{junction}) but with an 
extra factor of two. In the remainder we will work with 
equations (\ref{eom}) and (\ref{junction}) and we will refer to 
(\ref{junction}) as the junction condition.

The usual way \cite{RS} of establishing localization of gravity on the brane 
is to study small fluctuations around a  given configuration (such as a 
flat brane in AdS space) which solves equations (\ref{eom}). 
The equations for small gravitational fluctuations around the 
solution take the form of a quantum mechanical problem. In terms 
of the effective quantum mechanical problem 
the existence of a localized graviton translates into the existence
of a normalizable zero-mode solution
(this solution is the wave function associated to the 
graviton localized on the brane). In addition to the zero mode
there are additional massive modes. One still has to show 
that these modes do not drastically change the physics, 
i.e. that they yield subleading corrections relative to the zero mode.
Note that the question of normalizability of the zero mode
depends on global properties of the gravitational solution.
If the bulk space is asymptotically flat there is still 
a zero-mode but it is not normalizable. 
There may still be a quasi-localization due to a collection 
of low-energy Kaluza-Klein modes \cite{GRS,CEH,DGP}. 

The analysis just described is at the linearized level. 
It is technically involved in this approach to go beyond the linear 
approximation and demonstrate the full non-linear structure of the 
gravity localized on the brane. In this paper we use the AdS/CFT duality 
in order to achieve this goal. Previous works that use the AdS/CFT 
duality in the RS context include \cite{herman,Gub,gkr,HHR,DuffLiu,ANO,GK,DK}.

It has been shown in \cite{FeffermanGraham,HS}
that given a metric $g_{(0)}$ on the boundary of AdS 
one can obtain an asymptotic expansion of the bulk
metric near the boundary up to certain order in the radial 
coordinate (which is regarded as the small parameter in
the expansion). The next order coefficient is 
left undetemined by the bulk field equations \cite{FeffermanGraham}.
This coefficient is determined once a symmetric 
covariantly conserved tensor $T^{\tnnn{CFT}}_{ij}(x)$ with trace 
equal to the holographic Weyl anomaly is supplied.  
The tensor $T^{\tnnn{CFT}}_{ij}(x)$
is the holographic stress tensor of the dual conformal field 
theory \cite{KSS} (see also \cite{BFRS}). Notice that the CFT stress 
energy tensor encodes global information too. In particular, regularity 
of the bulk solution sometimes uniquely fixes $T^{\tnnn{CFT}}_{ij}(x)$.

Let us consider a brane placed close to the AdS boundary.
Then one can solve (\ref{eom}) by simply considering the 
asymptotic solution described in the previous paragraph.
The junction condition (\ref{junction}) then becomes 
Einstein's equation for the induced metric on the brane.
The right-hand side in Einstein's equations is equal to
the stress energy tensor due to matter localized on the brane plus 
the CFT stress energy tensor. 
In fact, irrespectively of the value of the bulk cosmological constant,
Einstein's equations in the bulk plus the junction condition 
effectively impose Einstein's equations on the brane.
This result first appeared in \cite{SMS}.
In particular in all cases the gravitational equations
on the brane involve a ``holographic stress energy tensor''.
This can be taken to holographically represent the bulk 
spacetime.

This paper is organized as follows. In the next section 
we adopt the results from the AdS/CFT duality to Brane-World
scenarios. In particular, we put a brane near the boundary of 
AdS and obtain the equation that the induced metric on the brane 
satisfies. In section 3 we place a brane at some (arbitrarily chosen)
position in the bulk and analyze the equations near the brane,
i.e. we consider the radial distance from the brane as 
a small parameter. These considerations are valid for
any bulk cosmological constant. Finally, in section 4
we consider the case of a brane placed near infinity
of an asymptotically flat bulk spacetime.

In this paper we only perform a local analysis. Global issues
are important and need to be addressed in order to establish 
localization of the graviton on the brane. This important 
issue is left for future study.

\section{Brane gravity from the asymptotic analysis of AdS space}

The asymptotic solutions of the bulk Einstein equation (\ref{eom}) in vacuum 
were worked out in \cite{HS} to sufficiently high order. These solutions are 
best found by writing the bulk metric in the Fefferman-Graham form 
\cite{FeffermanGraham}:
\be \label{GrFe}
\d s^2={l^2\over4\rho^2}\,\d\rho^2 
+{l^2\over\rho}\,g_{ij}(\rho,x)\d x^i\d x^j,
\ee
where the metric $g_{ij}$ has the expansion
\be\label{metric}
g(\rho,x)=g_{(0)}+\rho g_{(2)}+\cdots+\rho^{d/2}g_{(d)}+h_{(d)}\rho^{d/2}\log\rho+{\cal O}(\rho^{(d+1)/2}).
\ee
$l^2$ is related to the cosmological constant as $\L=-d(d-1)/2l^2$.
Given $g_{(0)}$ all coefficients up to $g_{(d)}$ can be found as 
local functions of $g_{(0)}$. The coefficient $g_{(d)}$ is undetermined from 
the gravity equations, and it is related to the stress-energy tensor of the 
dual CFT:
\be\label{T}
\< T_{ij}\>_{\tnnn{CFT}}={dl^{d-1}\over16\p G_{d+1}}\,g_{(d)ij} 
+X_{ij}^{(d)}[g_{(j)}],
\ee
where $X_{ij}^{(d)}[g_{(j)}]$ is a known function of the lower-order 
coefficients $g_{(j)}, j<d$ \cite{KSS} (see \cite{strings} for a review).
The gravitational equations imply that 
$\<T_{ij}\>_{\tnnn{CFT}}$ is covariantly conserved and its
trace reproduces the conformal anomaly of the boundary CFT. 

Let us place a brane close to infinity  at constant $\rho=\epsilon$, 
where $\epsilon$ is small enough for the expansion (\ref{metric}) to be 
a good approximation for the metric in the bulk.
Using the results of \cite{KSS}, it is now a simple matter 
(using (3.4),(3.5),(3.6) and (3.7) of \cite{KSS})
to see that the junction condition gives Einstein's equation on the brane.
For a 3-brane we get:
\bea\label{braneeq}
R_{ij}[\c]&-&{1\over2}\,\c_{ij}\,(R[\c] -{12 \over l^2}) 
+{1\over4} l^2 \log\e \left({1\over12}\nabla_i\nabla_iR[\c]
-{1\over4}\,\nabla^2R_{ij}[\c] +{1\over24}\, \c_{ij}\nabla^2R[\c]\right.\nonu
&+&\left.{1\over2}\,R^{kl}[\c]R_{ikjl}[\c] -{1\over6}\,R[\c]R_{ij}[\c] 
+{1\over24}\,\c_{ij}\,R^2[\c] -{1\over8}\,\c_{ij}\,R^{kl}[\c]R_{kl}[\c]\right)\nonu
&=&-16\p G_5 {1 \over l}\,
(\< T_{ij}[\c]\>_{\tnnn{CFT}} +T_{ij}^{\sm{bdry}}[\c]),
\eea 
where we kept only terms ${\cal O}(R^2)$, and there is an explicit 
dependence on the cutoff through the logarithmic term. 

There are several comments in order here:
\begin{itemize}
\item In deriving (\ref{braneeq}) it was essential that we added no 
counterterms to the action. Had we added counterterms, then all the 
curvature terms in the above formula would have been cancelled. Indeed, 
these precisely come from the infrared divergent part of the action.
\item In the effective Einstein equations the bulk spacetime
is represented by the holographic stress energy tensor. In other 
words, the Brane-World has a purely 
$d$-dimensional description where the bulk spacetime has been 
replaced by the cut-off CFT. The CFT couples to matter on the brane 
only through gravitational interactions. 
\item The effective Newton's constant is given by
\be \label{newton}
G_4 = {2 G_5 \over l}
\ee 
In the context of the two-sided RS scenario one should divide this 
result by two (see the discussion after (\ref{junction})).
\item The AdS/CFT duality predicts specific $R^2$-terms. The terms 
in (\ref{braneeq}) are derivable from the local 
action: $\int\d^dx\,a_{(4)}$, where $a_{(4)}$ is the 
holographic trace anomaly in four dimensions. 
\item The original expansion in the cutoff becomes an 
expansion in the brane curvature. 
\end{itemize}

It is straightforward to extend these results to higher
dimensions using the results in \cite{KSS}.

In $(2+1)$ dimensions the series in (\ref{metric}) 
terminates at the $\rho^2$-term, and one has the
exact expression \cite{KS}
\begin{equation}
g(x,\rho )=(g_{(0)}+{1\over 2}
g_{(2)}\rho )^2~,~~g_{(2)}={1\over 2}(R\,g_{(0)ij}+t_{ij})~~,
\label{2d}
\end{equation}
where $t_{ij}$ is conserved, $\nabla_{(0)}^j t_{ij}=0$, and its trace is
$\Tr\, t=-R$. It follows that $t_{ij}$ can be identified
as the Liouville stress energy tensor. 
The holographic stress energy tensor is equal to 
$\<T_{ij}\>={l \over 16 \p G_3}\, t_{ij}$.

Placing an one-brane at $\rho=\epsilon$ 
and neglecting $\epsilon^2$-terms one finds that 
the junction condition (\ref{junction}) implies
\begin{equation}
\gamma_{ij}=-8\pi G_3 (T^{\sm{bdry}}_{ij}+\<T_{ij}\>)~~,
\label{2dgrav}
\end{equation}
where $\gamma_{ij}={1\over \rho}g_{ij}(x, \rho )$ is the induced metric 
on the brane, and $T^{\sm{bdry}}_{ij}$ is the stress tensor 
of matter on the brane.
Note that in two dimensions there is no dynamical theory for just the metric 
tensor. Gravity induced on the one-brane is of the scalar-tensor type.

In the presence of matter in the bulk, it was shown in \cite{KSS} that the 
bulk equations can be solved in the same way. In this case, one again 
reinterprets the leading in $\epsilon$ terms as giving the terms in the 
action that determine the dynamics on the brane. For bulk scalar fields of mass $m^2=(\D-d) \D$, the effective brane action is:
\bea
S[\c,\F]&=&\int\,d^dx \sqrt{\c}\,\left[{1 \over 2(2 \D -d -2)}\F(x,\e) \Box_\c 
\F(x,\e) \right. \nonu
&&\left. 
+{(d-\D) \over 2} \left(1 +{1 \over 2 (d-1) (2 \D -d -2)} R[\c]\right) 
\F^2(x,\e) \right],
\eea
where again we only show the first few terms in the low energy  
expansion. The $d$-dimensional 
mass receives contributions both from the mass term in $(d+1)$ 
dimensions but also from the bending of the brane. Notice that 
a massless field in $d+1$ dimensions remains massless in $d$ dimensions.

\section{Local analysis}

In the previous section we made use of the asymptotic expansion of the bulk 
AdS metric (\ref{metric}). A similar analysis can be done for a brane 
located anywhere in the bulk by considering the local geometry near the brane.

Consider the Einstein equations in the bulk
\begin{equation}
R_{\mu\nu} + {2 \over d-1} \L G_{\mu\nu}=0.
\label{1.1}
\end{equation}
Near the brane one can use Gaussian normal
coordinates. In these coordinates the bulk metric takes 
the form 
\begin{equation}
\d s^2=\d r^2+\gamma_{ij}(r,x)\d x^i\d x^j,
\label{2}
\end{equation}
where  $r$ stands for the radial coordinate adjusted  so that the brane 
location is  at $r=0$.
Then the $(ij), (rr)$ and $(ri)$ components of 
Einstein equations (\ref{1.1}) read
\begin{equation}
R_{ij}[\gamma ]
+{2 \over d-1} \L \gamma_{ij}+{1\over 2}\partial^2_r\gamma_{ij}-{1\over 2}
(\partial_r\gamma\gamma^{-1}\partial_r\gamma)_{ij} 
+{1\over 4} \partial_r\gamma_{ij}\Tr(\gamma^{-1}\partial_r\gamma )=0
\label{3}
\end{equation}
\begin{equation}
{1\over 2}\partial_r (\Tr (\gamma^{-1}\partial_r \gamma ))
+{1\over4 }\Tr(\gamma^{-1}\partial_r \gamma )^2+
{2 \over d-1} \L =0
\label{4}
\end{equation}
\begin{equation}
\nabla_j [\gamma^{-1}\partial_r\gamma
- \Tr(\gamma^{-1}\partial_r\gamma )]^j_i=0.
\label{5}
\end{equation}
Combining the equations (\ref{3}) and (\ref{4}) we find that
\begin{equation}
R[\gamma ]+2 \L +{1\over 4}\left([\Tr(\gamma^{-1}\partial_r\gamma )]^2
-\Tr(\gamma^{-1}\partial_r\gamma )^2\right)=0.
\label{6}
\end{equation}

Let $\gamma_{ij}(x,r)$ have the following expansion near the brane: 
$$
\gamma=\gamma_{(0)}+\gamma_{(1)}r+\gamma_{(2)}r^2+...
$$
Then solving equations (\ref{3}), (\ref{4}) and (\ref{5}) 
iteratively we find expressions relating
the coefficients $\c_{(k)}$.  From equation (\ref{3}) we find that
\be
\Ric [\c_{(0)}]
+{2 \over d-1} \L
\c_{(0)}+\c_{(2)}-{1\over 2}\c^2_{(1)}+{1\over 4}\c_{(1)}\Tr\c_{(1)}=0.
\label{Ricci}
\ee
Equation (\ref{4}) to leading order gives 
\be
\Tr\c_{(2)}={1\over 4}\Tr \c^2_{(1)}-{2\over d-1}\Lambda.
\label{trace}
\ee
Taking the trace of (\ref{Ricci}) and using (\ref{trace}) 
one finds
\be
R[\c_{(0)}]+2\Lambda -{1\over 4}(\Tr \c^2_{(1)}-(\Tr \c_{(1)})^2)=0.
\label{Riccisc}
\ee
This equation can also be obtained from (\ref{6}).
Equation (\ref{4}) to the first two orders yields
\bea
&&\nabla^j\c_{(1)ij}=\nabla_i\Tr\,\c_{(1)}, \label{nabl1} \\
\nabla^j\c_{(2)ij} &=& \half \nabla_j
[\c_{(1)}^2-{1\over 2}\c_{(1)} 
\Tr \c_{(1)}-{1\over 4}\c_{(0)} (\Tr \c_{(1)}^2-(\Tr \c_{(1)})^2)]^j_i.
\label{divergences}
\end{eqnarray}
Equation (\ref{nabl1}) can be integrated as
\be \label{inte}
\c_{(1)} = t_{(1)} + \c_{(0)} \Tr\,\c_{(1)},
\ee
where $t_{(1)ij}$ is an ``integration constant'' that satisfies
$\nabla^i t_{(1)ij}=0$. 
One can check that (\ref{divergences}) is automatically satisfied
when (\ref{Ricci}) and (\ref{Riccisc}) are satisfied.

Forming the Einstein tensor, we obtain
\be
R_{ij}[\c_{(0)}]-{1\over 2}\c_{(0)ij}R[\c_{(0)}]
=\Lambda \c_{(0)ij}+T_{ij},
\label{Einstein1}
\ee
where 
\be
T_{ij}=-{2 \over d-1} \L \c_{(0)ij}-\c_{(2)ij} + \half \c_{(1)ij}^2 
-{1 \over 4} \c_{(1)ij} \Tr \c_{(1)} -{1 \over 8} \c_{(0)ij}
[\Tr \c_{(1)}^2 - (\Tr \c_{(1)})^2].
\ee
Equation (\ref{divergences}) implies that $T_{ij}$ is covariantly conserved.
In addition, equation (\ref{trace}) determines the trace of $T_{ij}$,
\be
\Tr\,T=
-2\Lambda -{(d-2)\over 8}\left(\Tr t_{(1)}^2-{1\over d-1}
(\Tr t_{(1)})^2\right)
\label{t2}
\ee

Let us now consider a physical brane with stress tensor 
$T^{\sm{bdry}}_{ij}$ located
at $r=0$. Then in addition to equations (\ref{3}), (\ref{4}), (\ref{5})
we have the  junction condition (\ref{junction}).
For the metric (\ref{2}) the second fundamental form is equal to 
$K_{ij}={1\over 2}\partial_r\gamma_{ij}$. From the junction 
condition (\ref{junction}) we get using the equation (\ref{nabl1})
\begin{equation}
t_{(1)ij}=16 \p G_{d+1} T^{\sm{bdry}}_{ij}.
\label{8}
\end{equation}
The junction condition thus identifies the 
undetermined covariantly conserved 
tensor $t_{(1)}$ in (\ref{inte}) with the stress tensor of the brane.
Notice that conservation of the boundary stress energy tensor 
is a neccessary condition for this identification.

To summarize we have shown that Einstein's equations in the 
bulk plus the junction condition lead to Einstein's equations 
on the brane. The effective stress energy tensor $T_{ij}$ 
represents both the bulk spacetime and the matter on the 
brane. Its trace is determined
by the matter stress energy tensor on the brane.
This is similar to the case discussed in the previous section.
There the effective stress energy tensor was a sum of 
the stress energy tensor of matter localized on the brane
of the $\<T_{ij} \>_{\sm{CFT}}$. The latter was taken to 
represent the bulk spacetime, and its trace was fixed to be the 
holographic conformal anomaly.
 
The results in this section agree with the results obtained 
in \cite{SMS} for $d=4$. To see this, let 
\be \label{SMS}
\c_{(2)ij}=-E_{ij} + {1 \over 4} \c^2_{(1)ij} -{2 \over d(d-1)} \L \c_{(0)ij},
\ee
and also let the boundary stress energy tensor be equal 
to  $T^{\sm{bdry}}_{ij}=-\l \c^{(0)}_{ij}+\t_{ij}$, where 
$\l$ is the tension and $\t_{ij}$ the matter energy momentum 
tensor on the brane. Equation (\ref{SMS}) defines the 
tensor $E_{ij}$. A short calculation shows that it agrees
with the tensor $E_{\m \n}$ of \cite{SMS}. In particular,
$E_{ij}$ is traceless and its divergence is equal to 
$\nabla^j E_{ij} = K^{jk}(\nabla_i K_{jk} - \nabla_j K_{ik})$.
This agrees with formula (22) of \cite{SMS}. 
One can also verify agreement with (17)-(20) of \cite{SMS}.

Note that the above considerations are
quite general and valid for any value of the bulk cosmological constant.
Note also that when the brane matter consists of only a brane 
cosmological constant the brane geometry has a constant Ricci scalar. 

\section{Asymptotically flat case}

In this section we perform an asymptotic analysis of Einstein's
equations with zero cosmological constant similar to the one 
that has been done for asymptotically AdS spaces in 
\cite{FeffermanGraham,HS}. 

We work in Gaussian normal coordinates. The metric takes the form
\be
\d s^2=\d r^2+\gamma_{ij}(x,r)\d x^i\d x^j.
\label{ds2}
\ee
Einstein's equations in this coordinate system are given in
equations (\ref{3}), (\ref{4}) and (\ref{5}). We look
for an asymptotic solution near infinity. Assuming
that the leading part of $\c$  near infinity is 
non-degenerate we find that it scales like $r^2$
(to prove this use  (\ref{3})). Restricting ourselves 
to this case, we look for solutions of the form
\be
\gamma (x,r)=r^2 (g_{(0)}+g_{(2)}{1\over r}+g_{(4)}{1\over r^2}+...)~~.
\label{asyF}
\ee
In other words, the bulk metric asymptots to a cone with $g_{(0)}$
the metric on the base. In general, one can include logarithmic 
terms in (\ref{asyF}). Such more general asymptotic solutions 
have been studied in \cite{BS,B}\footnote{In \cite{BS,B} the authors look for 
solutions whose metric coefficients near infinity is given by an 
expansion in negative powers of the radial coordinate. Coordinate 
transformations allow one to put the metric in  
the form $\d s^2=N^2 \d r^2+\gamma_{ij}(x,r)\d x^i\d x^j$,
with $N=1+\s(x)/r$ and $\gamma(x,r)$ as in (\ref{asyF}).
By a further logarithmic transformation one can reach 
Gaussian normal coordinates but at the expense of 
introducing logarithmic terms in $\gamma(x,r)$.
Our results for $d=3$ agree with the results of \cite{BS,B} for $\s(x)=0$.
We thank Kirill Krasnov for bringing these papers to our attention.}.
We restrict ourselves to (\ref{asyF}).
      
We solve Einstein's equations
\be
R_{\mu\nu}=0,
\ee
perturbatively in $1/r$.
The leading order equations imply \cite{GPP,SS} that $g_{(0)}$
should satisfy
\be \label{leading}
R_{(0)ij}+(d-1)g_{(0)ij}=0~~.
\ee
This means that the space at infinity is described by an Einstein metric
of constant positive scalar curvature. 
In particular, for Euclidean signature
the standard metric on the unit sphere $S^{d-1}$ satisfies
this equation. Then the leading part of the bulk metric (\ref{ds2}), 
(\ref{asyF}) is just 
Euclidean $R^d$ space. In the Lorentzian signature
equation (\ref{leading}) is solved by the de Sitter space.
Thus, already at leading order, we find an important difference 
between the cases of asymptotically flat spacetime and
of asymptotically AdS spacetimes. Whereas in the latter case one could 
choose the boundary metric arbitrarily, in the former case
the boundary metric has to satisfy (\ref{leading}).

To next order we find
\bea 
&&\nabla^jg_{(2)ij}=\nabla_i\Tr g_{(2)}, \label{na1} \\
&&d g_{(2)}+2 \Ric_{(2)}-g_{(0)}\Tr g_{(2)}=0, \label{Ei1}
\eea
where 
\be
\Ric[\c]=\Ric_{(0)}+{1\over r}\,\Ric_{(2)}+{1\over r^2}\,\Ric_{(4)}
\cdots
\ee
and
\be\label{R2ij}
R_{(2)ij}=-{1\over2}[\nabla_i\nabla_j\Tr g_{(2)}
-\nabla^2 g_{(2)ij} +2(d-1)g_{(2)ij}+2R_{(0)ikjl} g_{(2)}^{kl}],
\ee
where indices raised and lowered by $g_{(0)}$. In deriving 
this equation, (\ref{leading}) and (\ref{na1}) were used.
Then equation (\ref{Ei1}) becomes
\be
\nabla_i \nabla_j \Tr g_{(2)} - \nabla^2 g_{(2) ij} 
+(d-2) g_{(2)ij} + g_{(0)ij} \Tr g_{(2)} 
+ 2 R_{(0)ikjl}\, g_{(2)}^{kl}=0.
\ee
Notice that this equation leaves undetermined the trace of $g_{(2)}$.
Let us define
\be \label{deftij}
t_{ij} = g_{(2)ij} - g_{(0)ij} \Tr g_{(2)}.
\ee
It follows from the (\ref{na1}) that $\nabla^i t_{ij}=0$.

To the next order we find the equations
\bea \label{2order1}
g_{(4)}&=&-{1\over2}\Ric_{(4)} +{1\over2}g_{(0)} \Tr g_{(4)} 
-{1\over4} g_{(0)} \Tr g_{(2)}^2 +{1\over4}g_{(2)}^2 
+{1\over8}g_{(2)}\Tr g_{(2)}, \\
\Tr g_{(4)}&=&{1\over4}\Tr g_{(2)}^2, \label{2order2} \\
\nabla^j g_{(4)ij} &=& \half \nabla_j
[g_{(2)}^2-{1\over 2}g_{(2)} 
\Tr g_{(2)}-{1\over 4}g_{(0)} (\Tr g_{(2)}^2-(\Tr g_{(2)})^2)]^j_i,
\label{2order3}
\eea
and 
\bea
R_{(4)ij}&=&{1\over2}[-{1 \over 4} \nabla_i \nabla_j \Tr g_{(2)}^2 
-\nabla^k \nabla_i g_{(4)jk} -\nabla^k \nabla_j g_{(4)ik}
+ \nabla^2 g_{(4)ij} \nonu
&&+g_{(2)}^{kl} 
[\nabla_l \nabla_i g_{(2)jk} +\nabla_l \nabla_j g_{(2)ik} -\nabla_l
\nabla_k g_{(2)ij}] \nonu
&&+{1 \over 2} \nabla^k \Tr g_{(2)} 
(\nabla_i g_{(2)jk} +\nabla_j g_{(2)ik} - \nabla_k g_{(2)ij}) \nonu
&&+\half \nabla_i g_{(2)kl} \nabla_j g_{(2)}^{kl} 
+\nabla_k g_{(2)il} \nabla^l g_{(2)j}{}^{k}
-\nabla_k g_{(2)il} \nabla^k g_{(2)j}{}^{l}].
\eea
It may seem that by taking the trace of 
(\ref{2order1}) and using (\ref{2order2}) and (\ref{2order1})
one obtains a new equation for $g_{(2)}$. However, 
it turns out that the resulting equation is automatically 
satisfied. The same is true when taking the trace of (\ref{Ei1}) and using (\ref{R2ij}).

The equations we obtained look similar to the equations one gets in the 
case of asymptotically AdS spacetimes. There are important 
differences, however. In the case of asymptotically AdS spacetimes
the equations were algebraic, and they could be solved up to 
order $\r^d$. The coefficient $g_{(d)}$ was undetermined
except for its trace and divergence. In the case at hand the equations
for the coefficients are differential, and it is the trace 
of $g_{(2)}$ which is undetermined.

Let us place the brane at a fixed large radius $r=r_0 >> 1$. 
Then expanding the Einstein tensor 
for the induced metric $\gamma_{ij}$ we find that
\be
R_{ij}[\c]-{1\over2}\,\c_{ij}R[\c]=(d-2)\left({d-1\over2r_0^2}\,\c_{ij} 
+{1 \over 2 r_0}\,t_{ij}\right)+{\cal O}(1/r_0^2)~~,
\label{Einstein}
\ee
where $t_{ij}$ is given in (\ref{deftij}).
On the other hand we have
\be
K_{ij}-\gamma_{ij}K=-{d-1\over r_0}\gamma_{ij}-
{1 \over 2}t_{ij} + \co(1/r_0^2)~~.
\ee
Notice that this is the Brown-York stress
energy tensor \cite{BrownYork}. Thus, up to the leading divergence
in $r_0 \to \infty$, 
$t_{ij}$ is equal to the Brown-York stress energy tensor.
The junction condition on the brane gives a relation between 
$t_{ij}$ and the stress tensor $T^{\sm{bdry}}_{ij}$
of matter fields on the brane. Plugging back to (\ref{Einstein}) we find 
\be
R_{ij}[\gamma ]
-{1\over2}\gamma_{ij} R[\gamma ]
=-{(d-2)(d-1)\over2r_0^2}\,\gamma_{ij} 
-{(d-2)8\pi G_{d+1}\over r_0}\,T_{ij}^{\sm{bdry}}+{\cal O}(1/r_0^2),
\ee
i.e. we get Einstein's equations with negative cosmological 
constant $\Lambda=-{(d-1)(d-2)\over 2 r_0^2}$
and Newton's constant $G_{d}={(d-2) G_{d+1} \over r_0}$. The position
of the brane becomes the AdS radius of gravity on the brane.
Notice also that the formula for $G_d$ is the same with formula 
(\ref{newton}) with $l$ replaced by $r_0$.

{\bf Acknowledgements}: 
KS is supported in part by the NSF grant PHY-9802484.
SS is supported by the DFG Priority Programme SPP-1096. SdH would like to thank Soo-Jong Rey for hospitality and interesting discussions during his stay at Seoul National University.


\begin{thebibliography}{99}

\bibitem{RS} L. Randall and R. Sundrum, 
``An Alternative to Compactification'', 
Phys.Rev.Lett. {\bf 83} (1999) 4690-4693, hep-th/9906064.

\bibitem{jct} 
S.K. Blau, E.I Guendelman and A.H. Guth, 
``Dynamics of False-vacuum bubbles'', Phys. Rev {\bf D35} (1987) 1747.


\bibitem{GRS} R. Gregory, V. A. Rubakov, and S. M. Sibiryakov, 
``Opening Up Extra Dimensions
at Ultra-Large Scales,'' Phys. Rev. Lett. {\bf 84} (2000) 5928; 
hep-th/0002072.

\bibitem{CEH} C. Csaki, J. Erlich and T.J. Hollowood,
``Quasilocalization of gravity by resonant modes'', 
Phys.Rev.Lett. {\bf 84} (2000) 5932-5935, hep-th/0002161.

\bibitem{DGP} G. Dvali, G. Gabadadze and M. Porrati,
``Metastable gravitons and infinite volume extra dimensions'',
Phys.Lett. {\bf B484} (2000), 112-118, hep-th/0002190.

\bibitem{herman} H. Verlinde, ``Holography and Compactification'',
hep-th/9906182.

\bibitem{Gub} S.S. Gubser, ``AdS/CFT and gravity'', hep-th/9912001.

\bibitem{gkr}
S. B. Giddings, E. Katz, and L. Randall, ``Linearized Gravity in Brane
Backgrounds,'' JHEP {\bf 0003} (2000) 023; hep-th/0002091.

\bibitem{HHR}  S.W. Hawking, T. Hertog and H.S. Reall,
``Brane New World'', Phys.Rev. {\bf D62} (2000) 043501, hep-th/0003052.

\bibitem{DuffLiu} M.J. Duff and J.T. Liu,
``Complementarity of the Maldacena and Randall-Sundrum Pictures'',
Phys. Rev. Lett. {\bf 85} (2000) 2052, hep-th/0003237.

\bibitem{ANO} L. Anchordoqui, C. Nunez and K. Olsen,
``Quantum Cosmology and AdS/CFT'', JHEP {\bf 0010} (2000) 050,
hep-th/0007064.

\bibitem{GK} S.B. Giddings and E. Katz, 
``Effective theories and black hole production in warped compactifications'',
hep-th/0009176.

\bibitem{DK}  N.S. Deger and A. Kaya,
``AdS/CFT and Randall-Sundrum Model Without a Brane'',
hep-th/0010141.

\bibitem{FeffermanGraham}
{C. Fefferman and C. Robin Graham, `Conformal Invariants', in 
{\it Elie Cartan et les Math\'ematiques d'aujourd'hui} (Ast\'erisque, 1985) 
95.}

\bibitem{HS} M. Henningson and K. Skenderis, ``The holographic
Weyl anomaly'', JHEP {\bf 9807} (1998) 023, hep-th/9806087;
``Holography and the Weyl Anomaly'', hep-th/9812032.

\bibitem{KSS}
S. de Haro, K. Skenderis and S.N. Solodukhin,
``Holographic Reconstruction of Spacetime and Renormalization 
in the AdS/CFT Correspondence'', to appear in Commun. Math. Phys.,
hep-th/0002230.
 
\bibitem{BFRS} K. Bautier, F. Englert, M. Rooman and Ph. Spindel,
``The Fefferman-Graham Ambiguity and AdS Black Holes'',
Phys. Lett. {\bf B479} (2000) 291-298,
hep-th/0002156. 

\bibitem{SMS} T. Shiromizu, K. Maeda and M. Sasaki, 
``The Einstein Equations on the 3-Brane World'',
Phys.Rev. D62 (2000) 024012, gr-qc/9910076.

\bibitem{strings} K. Skenderis, ``Asymptotically Anti-de Sitter spacetimes
and their stress energy tensor'', hep-th/0010138.

\bibitem{KS} K. Skenderis and S.N. Solodukhin, 
``Quantum effective action from the AdS/CFT correspondence'',
Phys. Lett. {\bf B432} (2000) 316-322, hep-th/9910023. 

\bibitem{BS} R. Beig and B.G. Schmidt, ``Einstein's equations
near Spatial Infinity'', Commun.Math.Phys. {\bf 87} (1982) 65-80.

\bibitem{B} R. Beig, ``Integration of Einstein's equations near spatial
infinity'', Proc.R.Soc.Lond. {bf A 391} (1984) 295-304.

\bibitem{GPP} G.W. Gibbons, D.N. Page and C.N. Pope,
``Einstein metrics on $S^3$, $R^3$ and $R^4$ buddles'',
Commun.Math.Phys. {\bf 127} (1990) 529.

\bibitem{SS} S.N. Solodukhin, ``How to make the gravitational action on 
non-compact space finite'', Phys.Rev. {\bf D62} (2000), 044016, 
hep-th/9909197.

\bibitem{BrownYork} J.D. Brown and J.W. York,
``Quasilocal energy and conserved charges derived from the 
gravitational action'',
Phys.Rev. {\bf D47} (1993), 1407-1419. 

\end{thebibliography}
\end{document}